\documentclass[aps, pra, twocolumn,superscriptaddress,longbibliography]{revtex4-2}
\usepackage{amsmath}
\usepackage{graphicx}
\graphicspath{{./rydberg_figs/}}
\usepackage{dcolumn}
\usepackage{bm}
\usepackage{textcomp}
\usepackage{natbib}
\usepackage{textcomp}
\usepackage{upgreek}
\usepackage{enumerate}
\usepackage{multirow}
\usepackage{float}
\usepackage[
    colorlinks=true,
    linkcolor=blue,
    citecolor=blue,
    urlcolor=blue
]{hyperref}
\usepackage{xcolor}
\usepackage{ulem} 

\def\bra#1{\mathinner{\langle{#1}|}}
\def\ket#1{\mathinner{|{#1}\rangle}}

\begin{document}
\title{
Phase-Sensitive Heterodyne Detection of MW using EIT Harmonics\\ in Rydberg Atoms
}

\author{Mangesh Bhattarai}
\email{mbhatt22@jh.edu}
\affiliation{William H. Miller III Department of Physics and Astronomy, Johns Hopkins University, Baltimore, 21218, Maryland, United States}
\affiliation{Department of Physics, Indian Institute of Science, Bangalore, Karnataka, 560012, India}

\author{Rishav Hui}
\email{rishavhui@iisc.ac.in}
\affiliation{Department of Physics, Indian Institute of Science, Bangalore, Karnataka, 560012, India}

\author{Sumanta Khan}
\affiliation{School of Quantum Technology, Defence Institute of Advanced
Technology (Deemed University), Girinagar, Pune, Maharashtra, 411025, India.}
\affiliation{Department of Physics, Indian Institute of Science, Bangalore, Karnataka, 560012, India}

\author{Vineet Bharti}
\affiliation{Quantum Engineering Technology Labs, H. H. Wills Physics Laboratory and School of Electrical, Electronic, and Mechanical Engineering, University of Bristol, Bristol, BS8 1FD, United Kingdom}
\author{Vasant Natarajan}
\thanks{This paper is dedicated to the memory of Vasant Natarajan (1965--2021)}
\affiliation{Department of Physics, Indian Institute of Science, Bangalore, Karnataka, 560012, India}
\author{Kanhaiya Pandey}
\email{kanhaiyapandey@iitg.ac.in}
\affiliation{Department of Physics, Indian Institute of Technology Guwahati, Guwahati, Assam 781039, India}

\begin{abstract}

We investigate the generation and characterization of higher-order harmonics in the probe-laser absorption arising from nonlinear interactions in an electromagnetically induced transparency (EIT) ladder system involving Rydberg states and driven by two microwave fields in the heterodyne configuration. We characterize the amplitude and phase of the generated harmonics as a function of the relative frequency and phase of the applied microwave fields. The phase of the $n^{\mathrm{th}}$ harmonic follows the relation $\phi_n=n\phi$, demonstrating phase multiplication and suggesting that higher-order harmonics may offer an enhanced phase response for phase-sensitive measurements. We further characterize the harmonic amplitudes and bandwidths and find that the measured bandwidths are significantly larger than the intrinsic Rydberg-state linewidth, consistent with power broadening under the experimental conditions. The experimental observations are supported by density-matrix calculations, which reproduce the key features of the measured harmonic response.

\end{abstract}
	
	\maketitle	

\thispagestyle{empty}

\section{Introduction}\label{sec1}
Atomic-based quantum sensing has emerged as one of the most promising platforms for precision measurements and quantum technologies. Utilizing the well-defined and highly stable quantum properties of atoms, these sensors can achieve exceptionally high sensitivity and accuracy. Atomic quantum sensors have found applications in a wide range of fields, including atomic clocks~\cite{LBY2015, THF2005}, magnetometry~\cite{BUR2007}, gravimetry~\cite{MVP2018}, inertial sensing~\cite{GLM2020}, navigation~\cite{Garrido2019, WAM2022}, and tests of fundamental physics~\cite{HJH2015, AOK2020}.

In recent years, atomic systems have also attracted significant interest for microwave (MW) and radio-frequency (RF) sensing. Atomic-based MW sensors are based upon the phenomenon of electromagnetically induced transparency (EIT) involving Rydberg states driven by MW/RF field. The highly excited Rydberg states, owing to their large electric-dipole matrix elements, provide a promising platform for sensitive and broadband MW electric-field measurements \cite{KFK2017}. Such sensors can offer advantages such as wide frequency coverage and minimal perturbation of the field being measured, making them attractive for applications in wireless communication \cite{RYH2026}, electromagnetic-field metrology \cite{HSG2017}, radar \cite{WPC2025}, and quantum technologies \cite{APS2020}.

An important limitation of conventional Rydberg electrometry is that the atomic response primarily provides information about the field amplitude, frequency, and polarization \cite{SSK2013,FKS2015} whereas the phase of the electromagnetic wave contains additional information about propagation and spatial field structure. 
To overcome this limitation, phase-sensitive atomic interferometry schemes have been proposed. In particular, a closed-loop Rydberg system can establish quantum interference between multiple excitation pathways, converting the relative phase of microwave fields into an observable change in optical transmission. Such an approach enables simultaneous characterization of the microwave amplitude and phase and provides access to additional properties of the electromagnetic field, such as propagation direction and wavefront \cite{SNP2018}. Several studies have been done using such closed-loop systems \cite{BAR2023, ASG2022, KBW2025, MAP2020, MDP2017, BKN2021,ZSB2026}. 

The above method is based on interference among various excitation pathways, which depend on the phase. However, it is also possible to infer the phase from the interference of the two MW fields in space. A significant advance in phase-sensitive Rydberg sensing was the demonstration of an atomic superheterodyne receiver \cite{JHJ2020}. In this method, the phase of the unknown MW field is determined with respect to a reference microwave field. The two MW fields with different frequencies beat with each other, which is reflected as sinusoidal absorption behavior of the probe laser with respect to time. Recent studies of harmonic and intermodulation distortion have demonstrated that Rydberg atomic receivers exhibit nonlinear characteristics, including compression and harmonic generation originating from the nonlinear response of the atomic medium \cite{GZR2026,CQL2026}. 

In this work, we investigate the behavior of various harmonics for phase-sensitive microwave-field detection using Rydberg atoms with high MW fields. For high-power MW fields, the absorption behavior deviates from a sinusoidal nature due to higher harmonics being generated by the MW fields. We characterize the phase inferred through various harmonics. We also present the density matrix calculation to support the experimental observations. 
The paper is organized in the following way. In Section \ref{theory}, we discuss the theoretical framework for the experimental observation. Section \ref{setup}  describes the experimental setup. In Section \ref{Results} we present the experimental results, and discuss their implications. Finally, we present the conclusion of the paper in the last Section \ref{Conclusion}.         

\section{Theory}
\label{theory}
\begin{figure}
	\centering
	\includegraphics[width=0.48\textwidth]{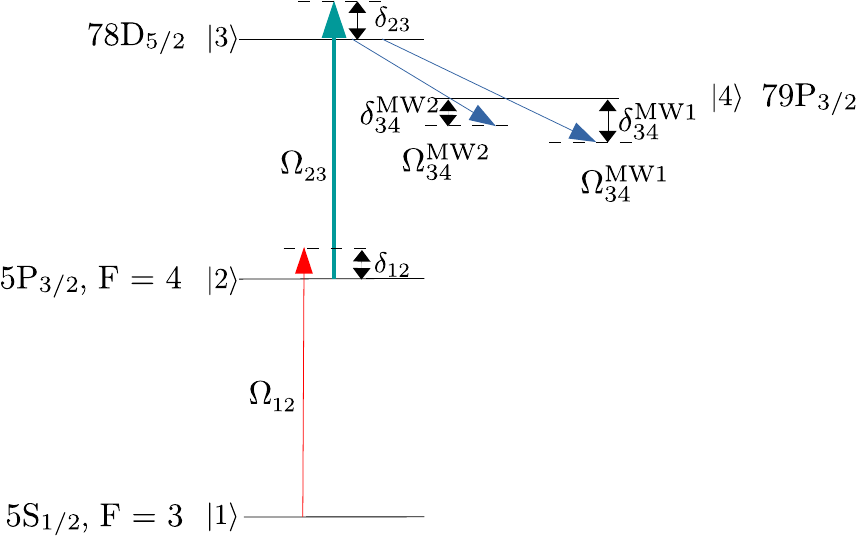}
	\caption{Energy level diagram showing all the relevant states and the couplings.}
	\label{fig:energy_level_theory}
\end{figure}
The relevant energy levels and the associated optical and microwave (MW) transitions are shown in Fig.~\ref{fig:energy_level_theory}. The probe and the control beams together form a simple ladder-type EIT system involving a Rydberg state. This Rydberg state in the ladder scheme is coupled to a nearby Rydberg state via two microwave fields.
The Rabi frequency of the probe laser driving the transition $\ket{1}\rightarrow\ket{2}$ is $\Omega_{12}$. The control laser drives  $\ket{2}\rightarrow\ket{3}$ with a Rabi frequency of $\Omega_{23}$. The Rabi frequencies of the two microwave fields, MW1 and MW2, that drive $\ket{3}\rightarrow\ket{4}$ are $\Omega^{MW1}_{34}$ and $\Omega^{MW2}_{34}$. The detunings of the lasers/MW fields are $\delta_{12}$, $\delta_{23}$, $\delta^{MW1}_{34}$ and $\delta^{MW2}_{34}$.
The Hamiltonian of the system in a rotating frame with the rotating wave approximation is given as 
\begin{align}
H=&-\delta_{12}\ket{2}\bra{2}-(\delta_{12}+\delta_{23})\ket{3}\bra{3}\\ \nonumber
&-(\delta_{12}+\delta_{23}-\delta^{MW1}_{34})\ket{4}\bra{4}\\ \nonumber
&+\Big[\frac{\Omega_{12}}{2}\ket{1}\bra{2}+\frac{\Omega_{23}}{2}\ket{2}\bra{3}\\ \nonumber
&+\frac{1}{2}\big\{\Omega^{MW1}_{34}+\Omega^{MW2}_{34}e^{i(\delta^{MW1}_{34}-\delta^{MW2}_{34})t}\}\ket{3}\bra{4}+h.c.\Big]
\end{align} 
Since the Hamiltonian is time dependent, the elements of the density matrix oscillate at multiple harmonics of $\Delta^{MW}=\delta^{MW1}_{34}-\delta^{MW2}_{34}$. 
The dynamics of the system can be described by the Lindblad master equation 
\begin{align}
\dot{\rho}=i[\rho, H]+ L\{\rho\}
\end{align}
where $L\{\rho\}$ is the Lindblad matrix representing the dissipation due to decay and decoherence~\cite{SNP2018}.

\begin{figure}
    \centering
    \includegraphics[width=0.45\textwidth]{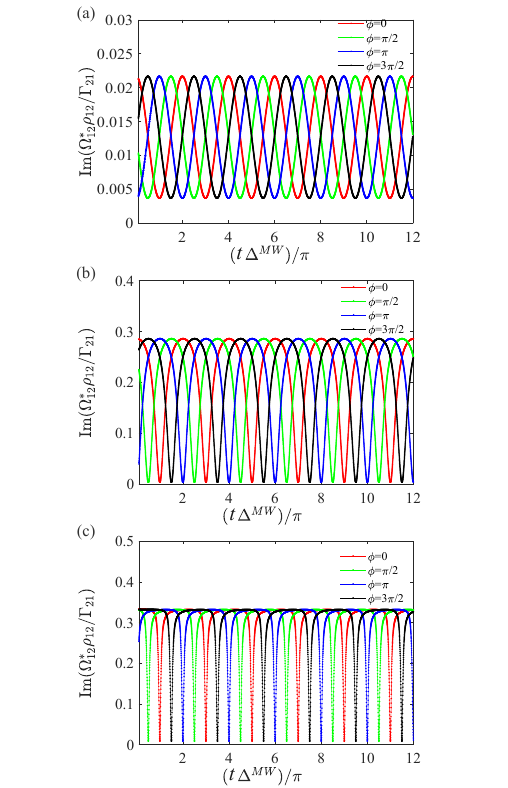}
    \caption{Probe absorption rate, Im($\Omega^*_{12}\rho_{12}/\Gamma_{21}$) vs time ($t\Delta^{MW}/\pi$) with $\Omega_{12}=\Gamma_{21}$; $\Omega_{23}=0.15\Gamma_{21}$, $\delta_{12}=\delta_{23}=\delta_{34}^{MW1}=0$. (a) $\Omega^{MW1}_{34}$ = $\Omega^{MW2}_{34}$ = 0.1$\Gamma_{21}$. (b) $\Omega^{MW1}_{34}$ = $\Omega^{MW2}_{34}$ = 1$\Gamma_{21}$. (c) $\Omega^{MW1}_{34}$ = $\Omega^{MW2}_{34}$ = 5$\Gamma_{21}$.} 
    \label{Abs_vs_time}
\end{figure}

\begin{figure}[htbp]
    \centering
    \includegraphics[width=0.45\textwidth]{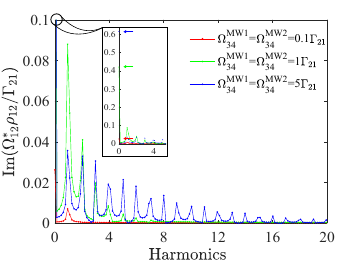}
    \caption{Harmonics of the probe absorption, Im($\Omega^*_{12}\rho_{12}/\Gamma_{21}$) for $\Omega_{12}=\Gamma_{21}$, $\Omega_{23}=0.15\Gamma_{21}$, $\delta_{12}=\delta_{23}=\delta_{34}^{MW1}=0$ with various MW Rabi frequency mentioned 
    on the labels. Inset shows larger Im($\Omega^*_{12}\rho_{12}/\Gamma_{21}$) range to accommodate the 0th-order harmonic; the arrows point to these values.}
    \label{Fourier_com}
\end{figure}

The probe absorption rate (normalized over the decay rate), defined as the imaginary part of ($\Omega_{12}^{*}\rho_{21}/\Gamma_{21}$), i.e., ($\mathrm{Im}[\Omega_{12}^{*}\rho_{21}/\Gamma_{21}]$), is plotted as a function of time (t), expressed in units of ($\pi/\Delta^{MW}$). For small values of ($\Omega_{34}^{MW1}$) and ($\Omega_{34}^{MW2}$), the probe absorption exhibits sinusoidal oscillations with a phase-dependent amplitude, as shown in Fig. \ref{Abs_vs_time}(a). As ($\Omega_{34}^{MW1}$) and ($\Omega_{34}^{MW2}$) are increased, the temporal variation of the probe absorption deviates from a sinusoidal form and evolves into a sequence of pulses separated by ($\Delta t\Delta^{MW}=2\pi$), as shown in Fig. \ref{Abs_vs_time}(b). This transformation arises from the generation of higher harmonics in the probe absorption. With a further increase in ($\Omega_{34}^{MW1}$) and ($\Omega_{34}^{MW2}$), the pulse width decreases as shown in Fig. \ref{Abs_vs_time}(c). This behavior is reminiscent of pulse formation in a mode-locked laser. The smaller width is due to increasing number of harmonics as shown in Fig. \ref{Fourier_com}.

To further analyze the various components of the harmonics, we do a Floquet expansion of the density matrix~\cite{S1965,NP2021} as below    
\begin{equation}
\rho=\sum^{\infty}_{m=-\infty}\rho^{(m)}e^{im\Delta^{MW} t}
\end{equation}

\begin{figure}[htbp]
    \centering
    \includegraphics[width=0.45\textwidth]{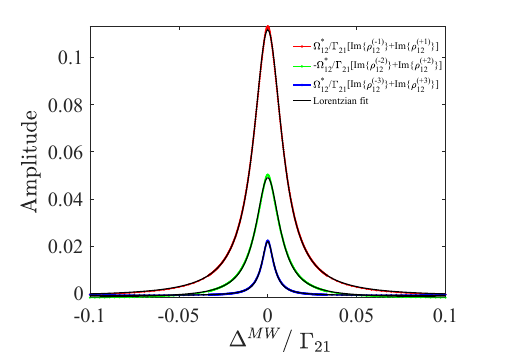}
    \caption{Harmonics of the probe absorption vs beat frequency $\Delta^{MW}/\Gamma_{21}$, for $\Omega_{12}=1.0\Gamma_{21}$, $\Omega_{23}=0.15\Gamma_{21}$, $\delta_{12}=\delta_{23}=\delta_{34}^{MW1}=0$ and $\Omega^{MW1}_{34}=\Omega^{MW2}_{34}=1\Gamma_{21}$. The linewidth after the Lorentzian fit is 0.02$\Gamma_{21}$, 0.02$\Gamma_{21}$ and 0.01$\Gamma_{21}$ for 1st, 2nd and 3rd harmonics respectively.}
    \label{bandwidth_theo}
\end{figure}

The quantity $\operatorname{Im}[\Omega_{12}\rho^{(\pm m)}_{12}/\Gamma_{21}]$ represents the probe absorption corresponding to the $m^{\mathrm{th}}$ harmonic. The amplitudes of the first three harmonics as a function of the microwave beat frequency, $\Delta^{MW}$, are shown in Fig.~\ref{bandwidth_theo}. The relatively large bandwidth with respect to $\Delta^{MW}$ can be attributed to power broadening.

\section{Experimental Setup}
\label{setup}
\begin{figure}
	\centering
	\includegraphics[width=0.45\textwidth]{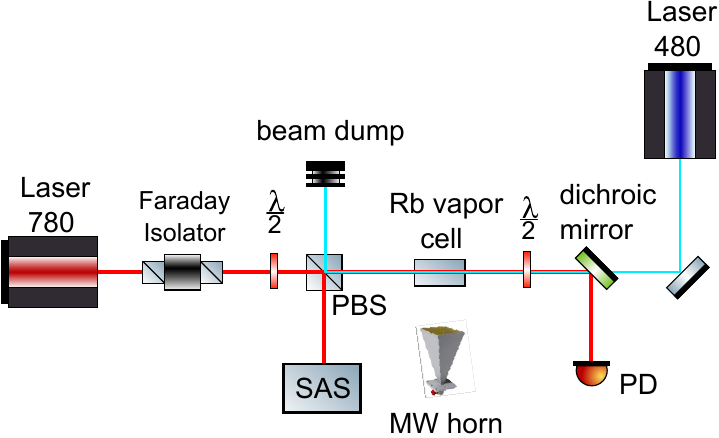}
	\caption{Schematic of the experimental setup. Figure key: PD--Photodiode, $\dfrac{\lambda}{2}$--Half Wave Retardation Plate, SAS--Saturated Absorption Spectrometer, PBS--Polarizing Beam Splitter.}
	\label{fig:setup}
\end{figure}

A schematic of the experimental setup is shown in Fig.~\ref{fig:setup}. A home-built external-cavity diode laser (ECDL) operating near 780 nm generates the probe beam of the ladder-type EIT system realized in the experiment. This laser is frequency-stabilized using the saturated absorption spectrum of the $5S_{1/2} ~F = 3 \to 5P_{3/2} ~F = 4$ transition in the D2 line of $^{85}$Rb. The probe beam has an elliptical beam shape with $1/e^2$ major and minor axes of 4.1 mm and 3.8 mm, respectively. 

Another laser, Toptica DL Pro, operating around 479.488 nm \cite{SPA2017} produces the control beam in the EIT scheme, that couples $5P_{3/2} ~F = 4$ to the Rydberg $78D_{5/2}$ state. The beam shape from this laser is elliptical with the $1/e^2$ major and minor axes of 446~\textmu m and 329~\textmu m, respectively. 

The probe and the control beam are guided into a room-temperature cylindrical Rb vapor cell (75 mm long and 25 mm in diameter) in a counter-propagating geometry as shown in the schematic. The power of the probe and the control beams entering the cell are about 100 \textmu W and 10 mW, respectively. As the probe beam exits the cell, it gets reflected at a dichroic mirror and is collected on a photodiode; the dichroic mirror, on the other hand, transmits the 480 nm laser beam into the vapor cell. A MW horn positioned a few centimeters from the cell illuminates the cell with MW field that propagates transverse to the optical beams, and is polarized perpendicular to the plane of the optical table, and parallel to the control beam polarization. The MW horn is designed to operate in the range $4 - 5$ GHz; in the experiment, it is used to transmit MW fields near 4.34 GHz. The microwaves are generated by SRS SG396 signal generators. 

\section{Results and discussion}
\label{Results}
\subsection{Characterization of EIT involving Rydberg states}

\begin{figure}
    \centering
    \includegraphics[scale=0.4]{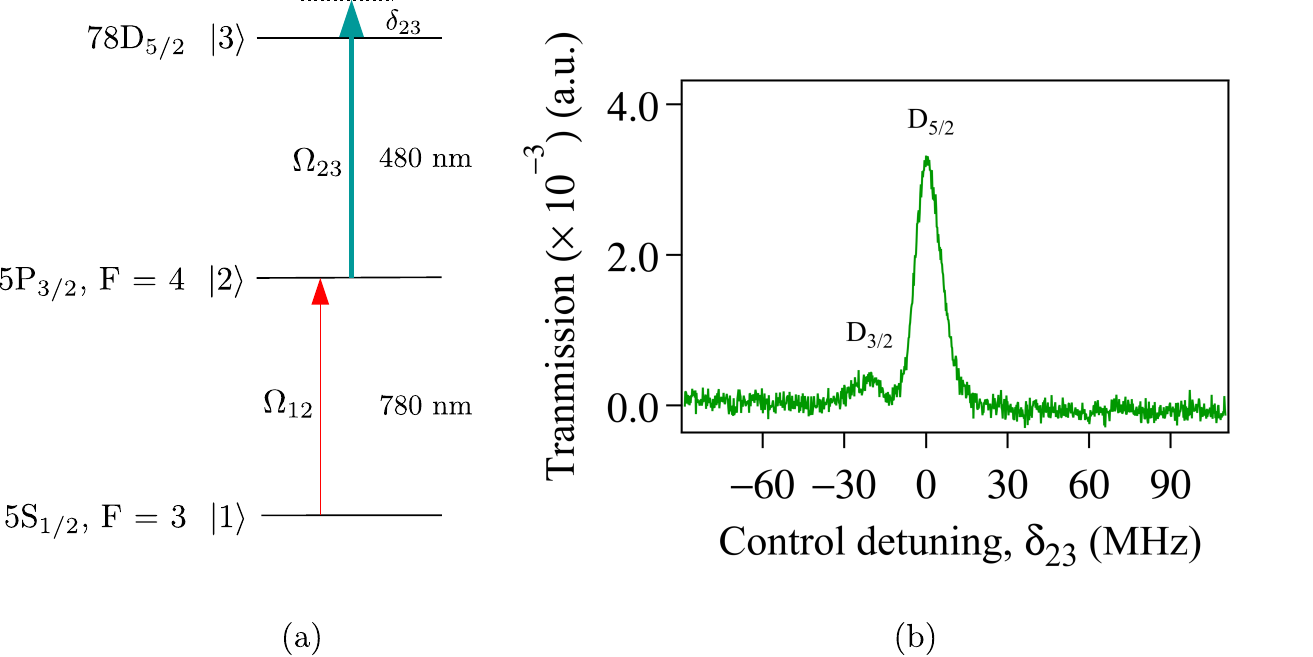}
    \caption{Rydberg EIT in a ladder-type system. (a) The relevant energy levels. (b) Probe transmission as a function of control beam detuning. The probe beam is locked to the $5S_{1/2} F = 3 \to 5P_{3/2} F=4$ transition, and the control beam is scanned around $5P_{3/2} F=4 \to 78D_{5/2}$ transition. Also visible in the trace at $\delta_{23} = -24$ MHz is the EIT peak corresponding to the closely lying $78D_{3/2}$ state. }
    \label{EIT}
\end{figure}

First, we characterize electromagnetically induced transparency (EIT) in the ladder-type system ($5S_{1/2}  F=3 \rightarrow 5P_{3/2}  F=4 \rightarrow 78D_{5/2}$), as shown in Fig.~\ref{EIT}(a). The probe laser, which drives the ($5S_{1/2}  F=3 \rightarrow 5P_{3/2}  F=4$) transition, is locked to resonance using saturated absorption spectroscopy. The control laser drives the ($5P_{3/2}  F=4 \rightarrow 78D_{5/2}$) Rydberg transition and is scanned across the resonance.

In the absence of a microwave (MW) field, we observe two EIT peaks corresponding to the closely spaced $78D_{5/2}$ and $78D_{3/2}$ Rydberg states, which are separated by approximately 24 MHz, as shown in Fig. \ref{EIT}(b).

\begin{figure}
	\centering
	\includegraphics[width=0.50\textwidth]{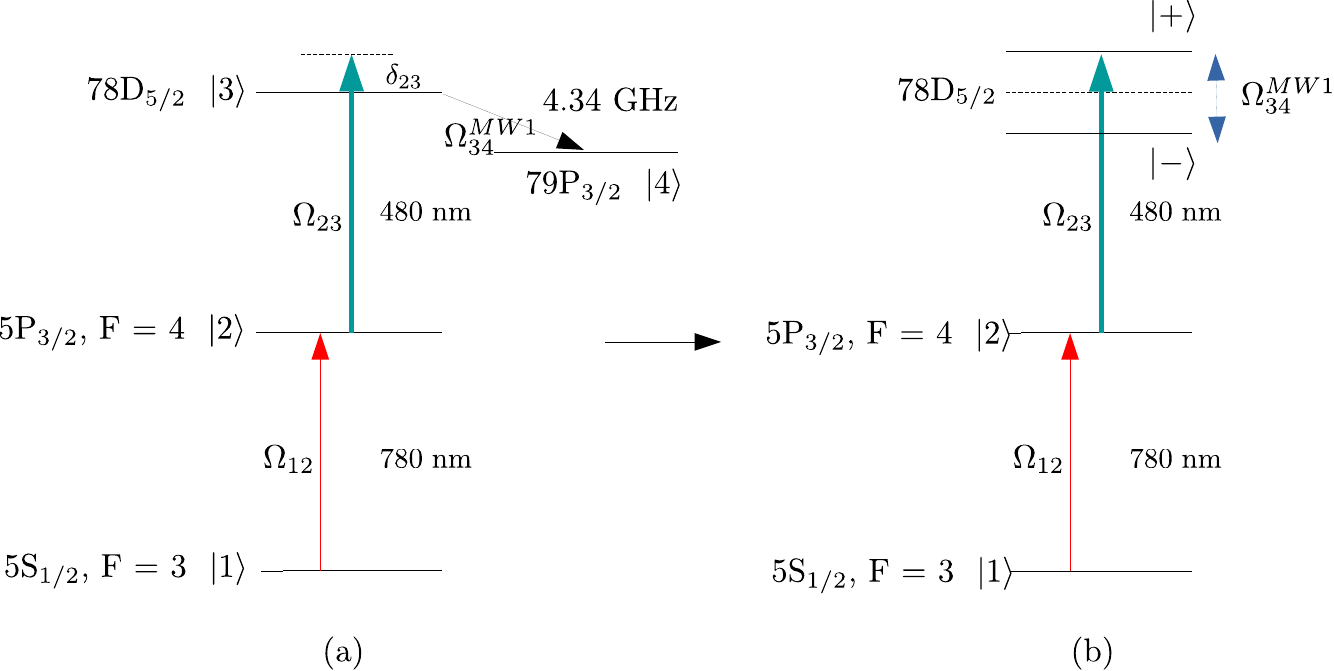}
	\caption{(a) Ladder EIT scheme with a MW at 4.34 GHz resonantly driving the transition between $78D_{5/2}$ and $79P_{3/2}$ with Rabi frequency $\Omega^{MW1}_{34}$. (b)
    MW coupling effectively creates dressed states $\ket{+}$ and $\ket{-}$ separated by $\Omega^{MW1}_{34}$ and positioned symmetrically about the mean energy of the $78D_{5/2}$ state.}
	\label{EITMW}
\end{figure}

\begin{figure}
	\centering
	\includegraphics[width=0.35\textwidth]{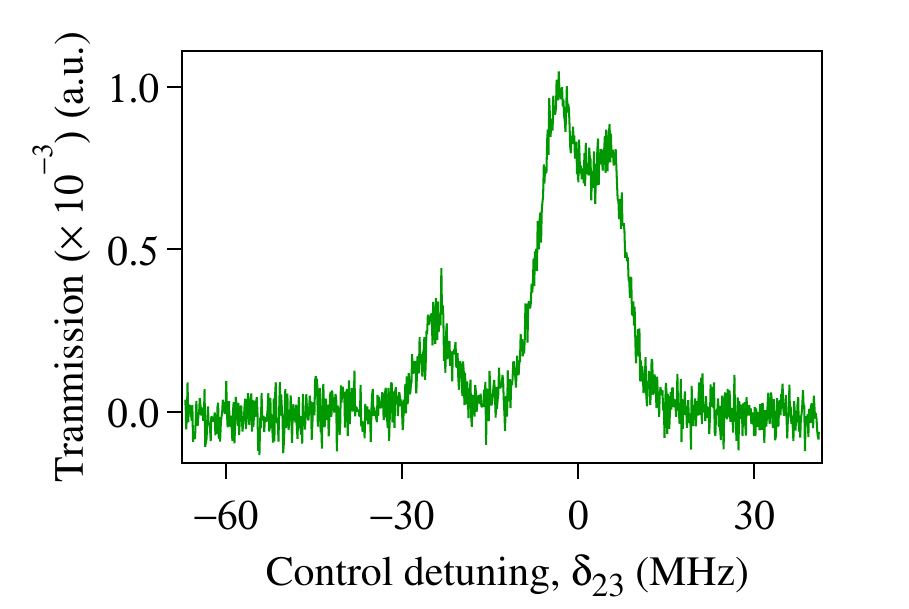}
	\caption{Splitting of the EIT signal at zero control detuning due to a resonant MW applied between $78D_{5/2}$ and $79P_{3/2}$. The power in the MW at the input of the horn is $-26$ dBm.}
	\label{EITMW_s}
\end{figure}

When a MW field at 4.34 GHz \cite{SPA2017} is applied to resonantly drive the ($78D_{5/2} \rightarrow 79P_{3/2}$) transition, as illustrated in Fig. \ref{EITMW}(a), the ($78D_{5/2}$) state is coupled to the ($79P_{3/2}$) state and forms two dressed states, denoted by ($\ket{+}$) and ($\ket{-}$), as shown in Fig. \ref{EITMW}(b). The energy separation between these dressed states is given by the MW Rabi frequency, ($\Omega^{MW1}_{34}$). Consequently, the dressing of the Rydberg states manifests as a splitting of the corresponding EIT resonance in the probe transmission spectrum, as shown in Fig. \ref{EITMW_s}.

\begin{figure}
	\centering
	\includegraphics[width=0.45\textwidth]{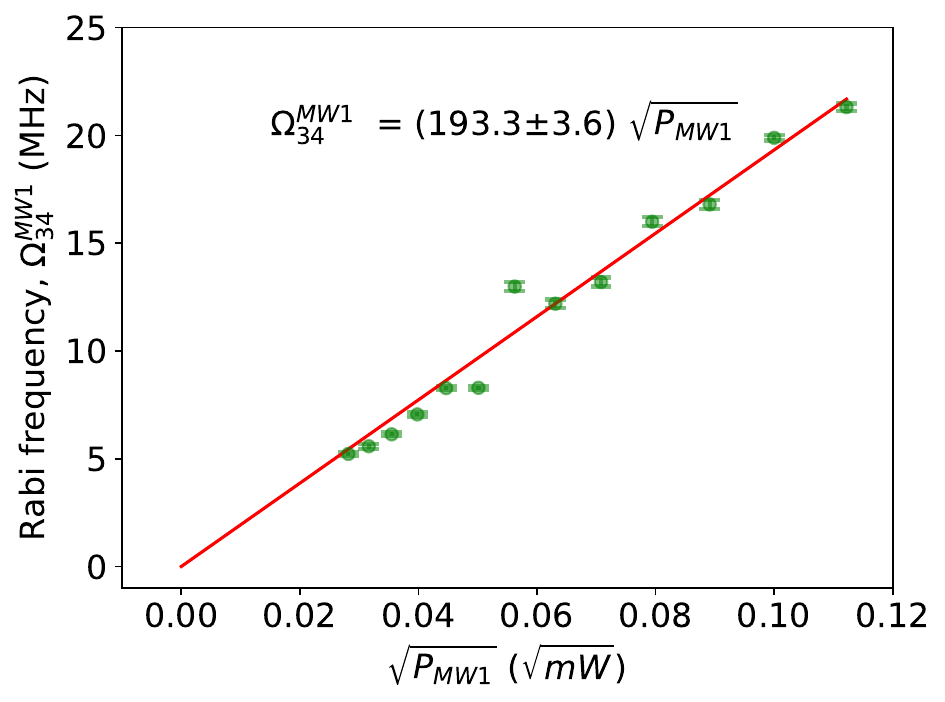}
	\caption{The Rabi frequency of the microwave field seen by the atoms as a function of square root of the power of MW field at the MW-horn input. The Rabi frequency is extracted by measuring the splitting of the EIT resonance induced by the MW. The MW resonantly couples the nearby Rydberg states indicated in Fig.~\ref{EITMW}(a). A linear fit, shown in red, captures the linear dependence of the MW Rabi frequency on the square root of the applied power of the MW field.}
	\label{EITMWspl}
\end{figure}

The observed splitting increases linearly with the MW electric-field amplitude, since the MW Rabi frequency is proportional to the field amplitude, ($\Omega^{MW1}_{34} \propto E_{MW1}$), and hence scales as the square root of the MW power, ($\Omega^{MW1}_{34} \propto \sqrt{P_{MW1}}$), as shown in Fig.~\ref{EITMWspl}. To determine the MW Rabi frequency, we fit the MW-split EIT spectrum with two Lorentzian profiles and compute the frequency separation between the two peaks. The MW Rabi frequency ($\Omega^{MW1}_{34}$), at the position of the beam, is extracted as $193.3$ MHz per $\sqrt{1~\rm mW}$ field strength at the horn input.


\subsection{Heterodyne detection of MW phases}
\begin{figure}
	\centering
	\includegraphics[width=0.45\textwidth]{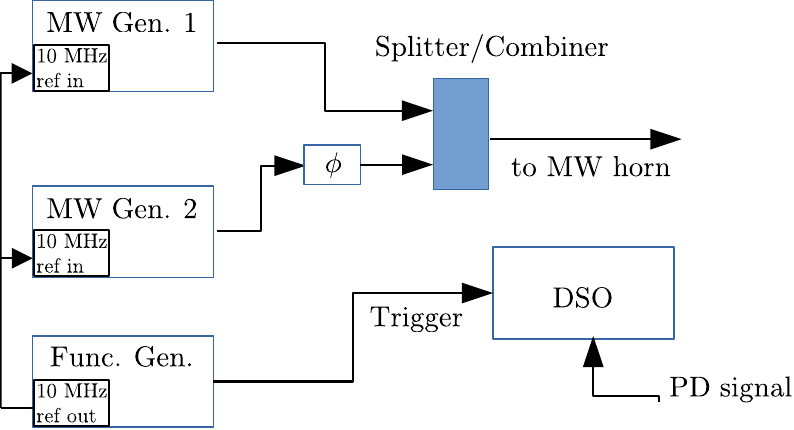}
	\caption{Schematic of the interference setup. Two independent MW sources---MW Gen 1 and MW Gen 2---operate near 4.34 GHz. These sources share a 10 MHz external clock reference, supplied by a function generator. The function generator is used to generate a square wave at the frequency difference of the two MW sources, which is used as a trigger to read the optical transmission of the probe beam on the oscilloscope (DSO). The phase of the MW signal from MW Gen. 2 is varied in the experiment, keeping the phase of MW Gen. 1 fixed.}
	\label{MWcomb}
\end{figure}

In the EIT setup described in the previous section, with the control beam at zero detuning from the EIT resonance with the $78D_{5/2}$ Rydberg state, we optically read out the interference between two independent microwave fields by monitoring the probe transmission under the EIT resonance condition.
The two MW fields are combined using a Mini-Circuits power combiner (ZN4PD-02183-S+), with one of the MW fields serving as the reference. The two MW signals are generated by independent microwave signal generators, MW Gen. 1 and MW Gen. 2, operating near 4.34 GHz as shown in Fig.~\ref{MWcomb}. Both signal generators are phase-locked to a common 10 MHz external frequency reference to ensure long-term frequency stability. The phase of MW Gen. 2 can be varied. The combined MW signal is then directed to the Rb vapor cell using the microwave horn.

A well-defined frequency difference is introduced between the two MW fields. To detect this time-dependent optical response, a function generator, whose internal oscillator provides the 10 MHz external reference clock, is used to generate a square-wave signal at the frequency difference between the two MW synthesizers. This signal serves as the trigger for recording the optical transmission of the probe beam. A schematic of the complete MW interference setup is shown in Fig. \ref{MWcomb}.

\begin{figure}
	\centering
	\includegraphics[width=0.5\textwidth]{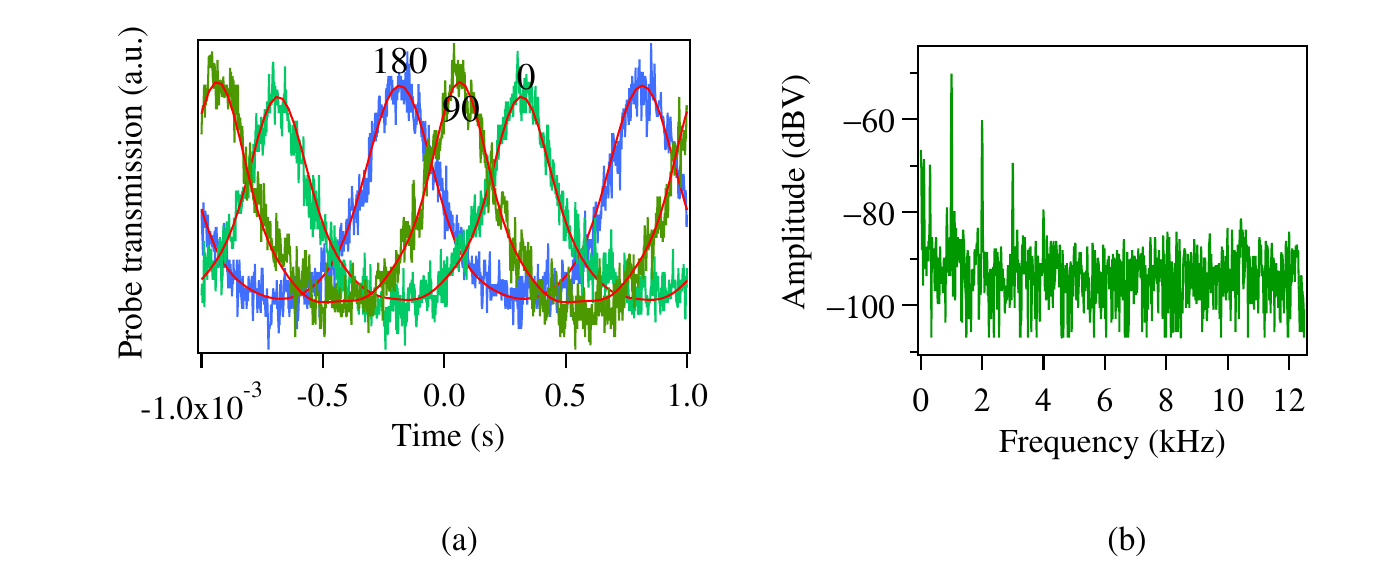}
	\caption{Interference between the two MW signals, read out by measuring the probe transmission. (a) Time domain signal. Numbers over each of the traces indicate the phase difference set at the MW Gen. 2. (b) FFT showing the different harmonics. $-30$ dBm power was sent from each MW Gen. into the horn.}
	\label{Intrf}
\end{figure}

The two superposed MW fields, emanating from the same microwave horn, produce a beat signal at their difference frequency, leading to a periodic modulation of the atomic response. As discussed in Sec. \ref{theory}, this modulation results in a time-dependent probe absorption, which is observed experimentally in Fig. \ref{Intrf}(a). As the phase difference between the two MWs is changed by changing the phase of the signal from MW Gen. 2, we see the interference signal read out with the probe, shifts in time as expected. Fig.~\ref{Intrf}(a) illustrates this for three different values of the phase difference set between the sources (at $0^{\circ}$, $90^{\circ}$, and $180^{\circ}$), where MW2 is detuned +1 kHz with respect to MW1. The interference signal in each case clearly shows evidence of deviation from a single-frequency sinusoidal behavior---the corresponding FFT signal in Fig.~\ref{Intrf}(b), observed on an oscilloscope, indicates the relative contribution of the different harmonics.

We record a series of such interference signals at different values of the phase difference between the two MWs from $0^{\circ}$ through $180^{\circ}$ in steps of $10^{\circ}$.
For the purpose of phase extraction, 
the time domain signal is fitted with a multi-sine function defined as follows:
\begin{equation}
    y = y_0 + \sum_{n=1}^3 a_n \sin(2\pi n f (t - t_0) +\phi_n) 
\end{equation}
with $a_n$, and $\phi_n$ as the free parameters, and $f$ fixed to the set frequency difference between the two MW sources. The amplitude $a_n$, then represents the strength of the interference signal in the $n$th harmonic.

\begin{figure}
	\centering
	\includegraphics[width=0.45\textwidth]{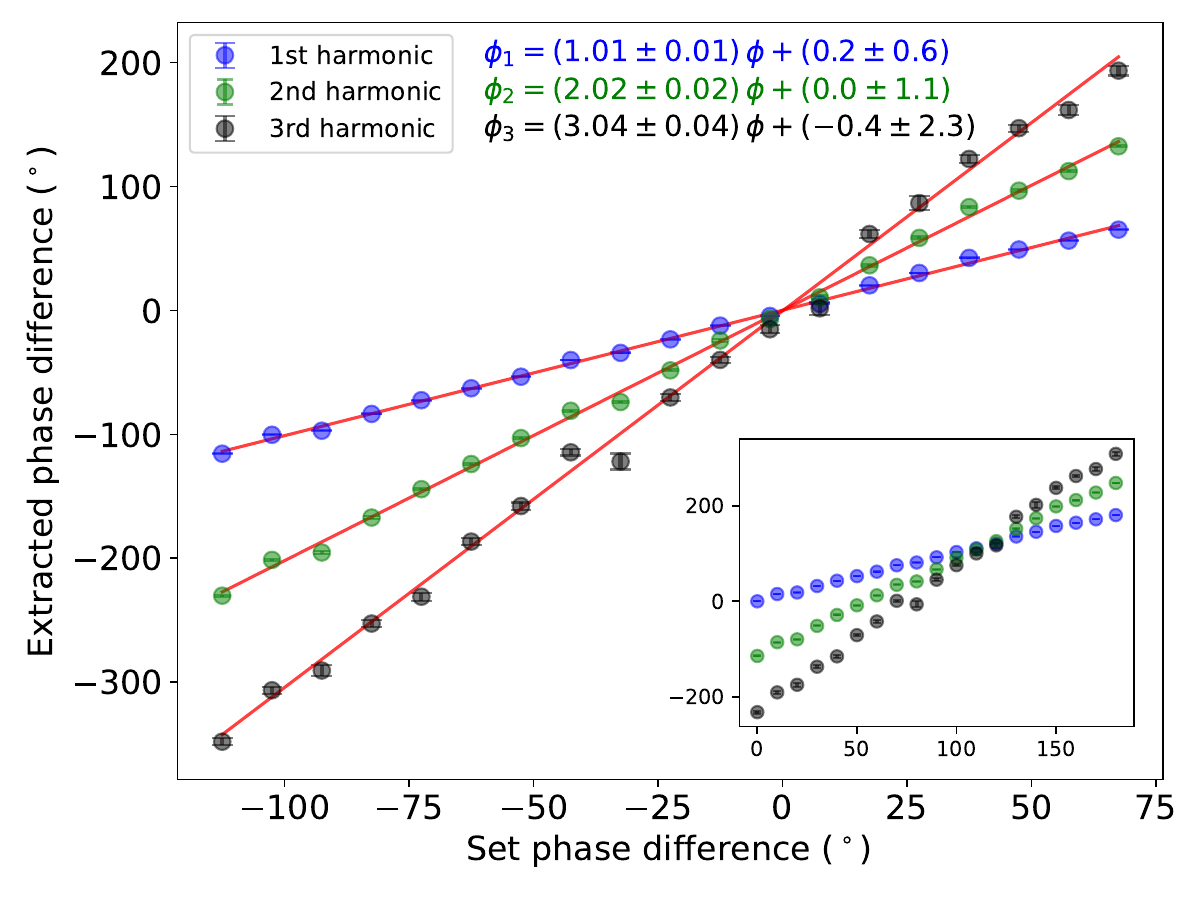}
	\caption{Phase, at different harmonics, extracted from a multi-sine function fit of the time-domain interference signal as a function of the applied phase difference. As detailed in the text, appropriate phase offsets have been subtracted from both, the phase difference set between the two microwave signal generator outputs, and the phase difference extracted from the fit. The $\phi_n = n \phi$ dependence of the extracted phase at $n$th harmonics on the actual phase difference between the two MW fields is evident from the linear fit shown in red for each group of harmonic signals. The error bars are extracted from the fit---a larger error bar signifies a lower SNR in the corresponding signal. The inset shows the values directly obtained from the fit. }
	\label{Phase}
\end{figure}

Initially, the multi-sine fit is performed with $t_0$ constrained such that the extracted phase of the first harmonic is zero at the set phase difference of zero. To ensure continuity in the extracted phase differences, the fitting routine uses the phase from the preceding fit as the initial point for the current fit. The extracted phase difference at different harmonics as a function of the set phase difference is shown in the inset of Fig.~\ref{Phase}. Since we expect the phase of all the extracted harmonics to be identically zero when the actual phase difference between the interfering MWs is zero, we add an appropriate offset in both the set and the extracted phase differences to realize this. The resulting dependence is shown in the main plot in Fig.~\ref{Phase} along with a linear fit to all the individual harmonic data depicting their $\phi_n = n\phi$ dependence, where $\phi_n$ is the phase difference read out on the $n$th harmonic signal and $\phi$ is the actual phase difference between the interfering MW fields. Consequently, the phase sensitivity increases linearly with the harmonic order, making higher-order harmonics particularly advantageous for precise phase determination.

\begin{figure}
	\centering
	\includegraphics[width=0.45\textwidth]{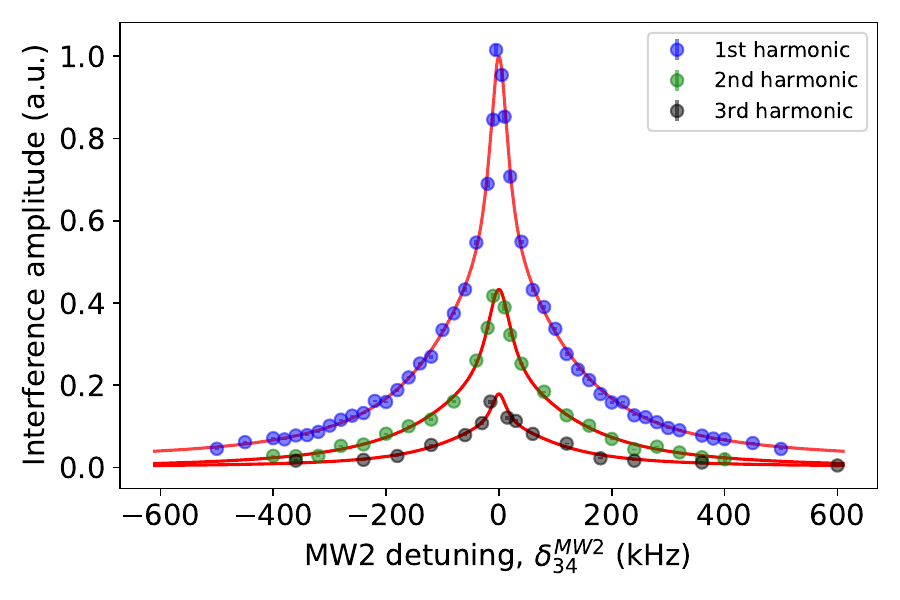}
	\caption{Interference amplitude for the 1st, 2nd and 3rd harmonics recorded against the detuning of MW2 from resonance. The amplitude has been rescaled to be unity at zero detuning for the 1st harmonic. The experimental data points are fit to a curve constructed as a sum of two Lorentzians with different amplitudes and linewidths but the same center, as defined in Eq.~\eqref{lor2}.  Good agreement between the experimental data points and the proposed curve suggests the existence of two different broadening phenomena characterized by different decay rates.  }
	\label{Bandwth}
\end{figure}

We next investigate the amplitude response of the different harmonics as a function of the frequency difference between the two MW fields, as shown in Fig. \ref{Bandwth}. For this, the output of both MW Gen. 1 and MW Gen. 2 is set to be $-26$ dBm at the horn input, and the frequency of the MW Gen 2 is varied $\pm 600$ kHz around the resonance frequency. The probe transmission, at each value of MW Gen. 2 detuning, is recorded on a Tektronix RSA 5103B spectrum analyzer. The recorded spectrum contains several harmonics at the detuning frequency, qualitatively similar to the FFT trace in Fig.~\ref{Intrf}(b). Each harmonic is fitted with a Gaussian curve, and the extracted amplitude is used to create the frequency response of the atomic system.

The characteristic bandwidth measured in the experiment is significantly larger than the intrinsic linewidth of the Rydberg states, primarily due to power broadening, as discussed in Sec.~\ref{theory}.   
More interestingly, a function of the following form best fits the data for each harmonic.
\begin{equation}
    y(f) = y_0 + a_1 \dfrac{\gamma_1^2/4 }{f^2 + \gamma_1^2/4 } 
    +a_2 \dfrac{\gamma_2^2/4 }{f^2 + \gamma_2^2/4 }
\label{lor2}
\end{equation}
The fit parameters are summarized in Table~\ref{table}.
A very good agreement between the experimental data and the proposed curve indicates the existence of two different broadening phenomena contributing to the overall response of the atoms---the decay rates associated with both phenomena are significantly larger than the decay rate of the Rydberg states involved. The identification of the process leading to such characteristic response is beyond the scope of the current study; we plan to identify and analyze it in detail in future work.

\begin{table}[htbp]
    \centering
    \caption{Fit parameters for the first three harmonics.}
    \label{tab:harmonic_fit_parameters}
    \begin{tabular}{lrrrr}
        \hline\hline
        Harmonic &
        $a_1$ &
        $a_2$ &
        $\gamma_1\;(\mathrm{kHz})$ &
        $\gamma_2\;(\mathrm{kHz})$ \\
        \hline
        1st &
        $0.55$&
        $0.43$&
        $40.6$ &
        $284.5$ \\

        2nd &
        $0.24$&
        $0.19$&
        $53.1$&
        $307.6$\\

        3rd &
        $0.08$&
        $0.09$&
        $34.0$ &
        $240.0$ \\
        \hline\hline
    \end{tabular}
    \label{table}
\end{table}




\subsection{Microwave four-wave mixing}

\begin{figure}
	\centering
	\includegraphics[width=0.45\textwidth]{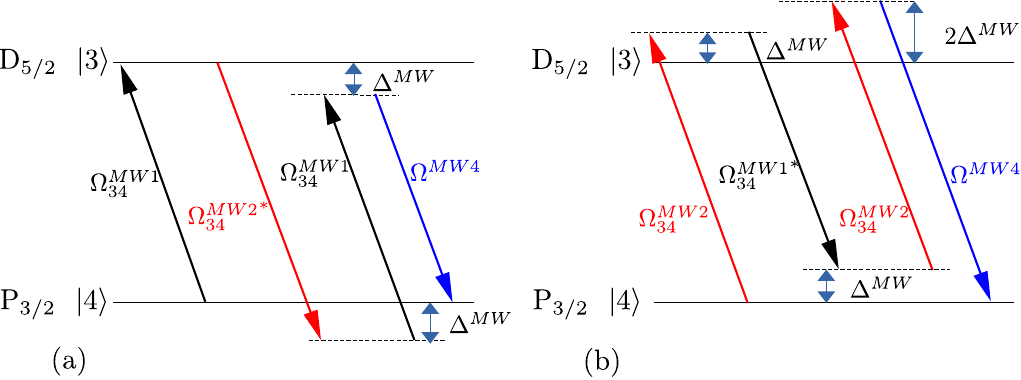}
	\caption{Two processes giving rise to second harmonics. The origin of the harmonics can be attributed to the four-wave mixing process occurring solely within the $D_{5/2}$ and the $P_{3/2}$ Rydberg states.}
	\label{fig:wavemixing}
\end{figure}

The origin of the higher harmonic response can be attributed to four-wave and higher-order wave mixing. 
Fig.~\ref{fig:wavemixing} summarizes two processes that contribute to the origin of the second harmonic response. We describe the process within the effective two-level system coupled by the two MW fields. 
Fig.~\ref{fig:wavemixing}(a) depicts a process which involves---from left to right---absorption of MW1, followed by stimulated emission of MW2, followed by absorption of MW1 again, at the end of which a four-wave mixed microwave is emitted, here represented as MW4. The Rabi frequencies of each of the fields are written alongside the arrow indicating the transition. Throughout the interference characterization, we have fixed MW1 at resonance and fixed the phase at $\phi_1$, while the frequency and the phase of MW2 are varied; the relative frequency of MW2 is represented as $\delta_{34}^{MW2} = \Delta^{MW}$, and its phase as $\phi_2$. We have,  $\Omega_{34}^{MW1} = |\Omega_{34}^{MW1}|e^{i\phi_1}$, and $\Omega_{34}^{MW2} = |\Omega_{34}^{MW2}|e^{i\phi_2} e^{-i\Delta^{MW}t}$. The Rabi frequency of the four-wave mixed MW is \cite{DJM2007}:
\begin{align}
    \Omega^{MW4} & \propto \Omega_{34}^{MW1}{\Omega_{34}^{MW2}}^*\Omega_{34}^{MW1} \\ \nonumber
                & = \left|\Omega_{34}^{MW1}\Omega_{34}^{MW2}\Omega_{34}^{MW1}\right| e^{i\Delta^{MW}t} e^{i(2\phi_1-\phi_2)}
\end{align}
This MW field is generated within the atomic medium, at a frequency $\omega_0 - \Delta^{MW}$ ($\omega_0$ being the frequency separation between the two states) and a phase of $(2\phi_1 - \phi_2)$. It can interfere with the MW2, at frequency $\omega_0 + \delta^{MW}$, and phase $\phi_2$, generating modulations proportional to $\cos (2\Delta^{MW2}t + 2\Delta\phi)$, where $\Delta\phi = \phi_1-\phi_2$, which in part contributes to the observed interference signal at the second harmonic of the frequency difference.

A similar analysis of the process outlined in Fig.~\ref{fig:wavemixing}(b) produces a four-wave mixed MW field at a frequency of $\omega_0 + 2\Delta^{MW2}$ and at a phase $2\phi_2-\phi_1$. This field can interfere with MW1, and thus produce an interference that goes as  $\cos (-2\Delta^{MW2}t - 2\Delta\phi)$. It is worth noting that these two processes occur in-phase and lead to enhancement of the second harmonic signal.

This analysis can be easily extended to describe the generation of the third and subsequent harmonics.

\vspace{5mm}
\section{Conclusion}
\label{Conclusion}

The results presented here demonstrate that the nonlinear response of a Rydberg-EIT system can be used to generate and resolve multiple harmonics of microwave-induced modulation in the probe absorption.
The observed phase multiplication with harmonic order provides a direct connection between the phase of the applied microwave field and that of the generated harmonics.
This could be used for improved phase sensitivity of the MW field.
The measured harmonic spectra also reveal bandwidths well beyond the intrinsic Rydberg-state linewidth due to power broadening.
Both the origin of the generated harmonics and their phase and spectral characteristics are consistently captured by the density-matrix model.
These results establish nonlinear Rydberg-EIT as a useful platform for studying coherent microwave interactions and motivate further investigation of higher-order harmonics for precision microwave field characterization.


\section*{Acknowledgments}

We thank Vibhor Singh for lending us the SG396 signal generators and the Tektronix spectrum analyzer, and S Raghuveer for valuable assistance with component procurement.
K.P. acknowledges funding from DST, India through Grant No. DST/QTC/NQM/QC/2024/1 (G).

\section*{Data availability} Data will be provided by the authors on a reasonable request.

\end{document}